\documentclass[reqno]{amsart} \usepackage{amscd}
\usepackage{epsfig}

\theoremstyle{definition}

\theoremstyle{remark} 
\numberwithin{equation}{section}
\newcommand{\be}{\begin{eqnarray}} 
\newcommand{\ee}{\end{eqnarray}}
\newcommand{\field}[1]{\ensuremath{\mathbb{#1}}}
\newcommand{\CC}{\field{C}}

\DeclareMathOperator{\SL}{SL}

\DeclareMathOperator{\SU}{SU}

\begin{document}

\title[Racah Coefficients]{Holography for the Lorentz Group \\ Racah Coefficients}
\author{Kirill Krasnov} \address{School of Mathematical Sciences \\
  University of Nottingham \\ Nottingham, NG7 2RD, UK and
Bogolyubov Institute for Theoretical Physics \\ Metrologichna 14 b \\
  Kiev, 03143, Ukraine}
\email{krasnov@maths.nott.ac.uk}
\begin{abstract} A known realization of the Lorentz group Racah coefficients is given
by an integral of a product of 6 ``propagators'' over 4 copies of the hyperbolic space. These
are ``bulk-to-bulk'' propagators in that they are functions of two points in the hyperbolic space.
It is known that the bulk-to-bulk propagator can be constructed out of two bulk-to-boundary ones. We point out
that there is another way to obtain the same object. Namely, one can use
two bulk-to-boundary and one boundary-to-boundary propagator. Starting from this construction
and carrying out the bulk integrals we obtain a realization of the Racah coefficients that
is ``holographic'' in the sense that it only involves boundary objects. 
This holographic realization admits a geometric interpretation in terms of an
``extended'' tetrahedron.
\end{abstract}
\maketitle

\section{Introduction}
\label{sec:intr}

Racah coefficients or, as they are often called, $6j$-symbols play the key role in the state sum approach to
quantum gravity in 2+1 dimensions. This approach, pioneered by paper \cite{PR}, has recently been revived
under the name of spin foam models. To get a spin foam model one triangulates the manifold in question, labels edges 
of the triangulation by the irreducible representations of some gauge group, associates to every tetrahedron
the $6j$-symbol constructed out of the corresponding representations, multiples the $6j$-symbols and sums
over the representation labels. The resulting ``partition function'' is triangulation independent and gives
a topological invariant of the manifold. The original model \cite{PR} of Ponzano and Regge uses $\SU(2)$ as
the gauge group. The Ponzano-Regge model has an interpretation of computing the path integral of Euclidean signature 3d 
gravity with zero cosmological constant. Other choices of signature and/or cosmological constant lead to
other gauge groups. In particular, the introduction of a cosmological constant leads to appearance of quantum groups,
with the deformation parameter being a function of the cosmological constant. 

Particularly important for physical applications is the case of negative
cosmological constant. It is widely believed that in this case (and when the signature is Euclidean) 
the relevant group is some
quantization of Lorentz group $\SL(2,\CC)$. ``Some quantization''
refers to the known fact \cite{Wor} that there exist inequivalent quantum deformations of 
$\SL(2,\CC)$. It is expected that there is a model analogous to that of Ponzano-Regge that
``solves'' the Euclidean 3d quantum gravity with negative cosmological constant and whose
building blocks are $6j$-symbols of an appropriate quantum group $\SL_q(2,\CC)$. 

In this paper, as a step in the direction of finding the corresponding spin foam model, 
we consider $6j$-symbols of the classical Lorentz group. We derive a certain formula
for these quantities that is given by a set of integrals over the boundary of the 
hyperbolic space. The expression we obtain is ``holographic'' in the sense that
only the boundary objects are involved. Thus, even though the object of
eventual interest for physical applications is the Racah coefficient of a quantum $\SL(2,\CC)$,
we shall study its classical counterpart hoping that an analog of the realization we obtain
also exists in the quantum case. 

These remarks being made, let us remind the reader some key facts about the Lorentz group 
representation theory. We shall be brief; the reader can consult e.g. \cite{Gelfand} for more details.
Representations of the Lorentz group
$\SL(2,\CC)$ can be realized in the space of homogeneous functions on 
the light cone in Minkowski space $M^{1,3}$. Representations of the type I,
which are of prime interest for physical applications, can also be realized in the space
of square integrable functions $L^2(H_3)$ on the hyperbolic space $H_3$. We define the 
Racah coefficients using this $L^2(H_3)$ realization
of the representations. 

Let us introduce some convenient terminology. Minus degree of 
homogeneity will be referred to as the conformal dimension $\Delta$ of the representation. For representations
of the type I:
\be\label{type-1}
\Delta=1 + i \rho.
\ee
We shall also require a notion of the dual representation. Its conformal dimension
$\bar{\Delta}$ is such that:
\be\label{sum}
\Delta+\bar{\Delta}=2.
\ee
The dual representation is an equivalent representation. 

The Racah coefficients  will be defined as a certain integral over several copies of
$H_{3}$. To write down the corresponding expression we need notions of bulk-to-bulk and bulk-to-boundary
propagators (the terminology is borrowed from literature on AdS/CFT correspondence). Let us consider the
upper half-space model of $H_3$. Let $(\xi_0>0,\xi), \xi\in\CC$ be the coordinates in the upper half-space. 
The $H_3$ metric is given by:
\be
ds^2 = \frac{1}{\xi_0^2}(d\xi_0^2 + |d\xi|^2).
\ee
The boundary of $H_3$ is the set of points $\xi_0=0$. When referring to boundary points we shall use
a different letter $x: x=\xi$.

Let us now introduce the so-called bulk-to-boundary propagator. This object has re-appeared in the
physics literature in the context of AdS/CFT correspondence, see \cite{Witten}. In the mathematics
literature it goes under the name of the kernel of the Gelfand-Graev transform, see \cite{Gelfand}. 
As the name bulk-to-boundary propagator suggests, it is a function of a boundary point $x$  
and a bulk point $\xi$:
\be\label{bound-prop}
K_\Delta(\xi,x) = \frac{\xi_0^\Delta}{(\xi_0^2+|\xi-x|^2)^\Delta}.
\ee
Another, more familiar from mathematics literature, expression for this object is given by:
\be\label{bound-prop-1}
K_\Delta(\xi,x)=(x\cdot \xi)^{-\Delta},
\ee
where $x,\xi$ are vectors in Minkowski space $M^{1,3}$ that correspond to points $x,\xi\in H_3$ 
in the hyperboloid model of $H_3$, and $(\xi\cdot\eta)$ is the usual Minkowski space bilinear pairing.

The so-called bulk-to-bulk propagator can be obtained by taking a product of two propagators 
\eqref{bound-prop}, one for the conformal dimension $\Delta$ and another one for the dual conformal dimension $\bar{\Delta}$,
and integrating over the boundary point:
\be\label{bulk-prop}
K_\Delta(\xi_1,\xi_2)= \int_{S^2} d^2x K_\Delta(\xi_1,x) K_{\bar{\Delta}}(\xi_2,x).
\ee
This way of getting the bulk-to-bulk propagator has been used in e.g. \cite{BC}.
The bulk-to-bulk propagator carries an orientation. Change of orientation replaces the representation
with its dual. Both the bulk-to-bulk and bulk-to-boundary propagators exist for an arbitrary value of $\Delta$,
not just for those \eqref{type-1} corresponding to representations of type I.

We are now ready to give a formula for the Racah coefficients. We shall give an expression for a general set of
conformal dimensions $\Delta_1,\ldots,\Delta_6$, and then discuss the conditions that $\Delta_i$ should satisfy
in order for this expression to make sense. We also refer to the Racah coefficients as $(6\Delta)$ symbols.
Following \cite{BC,FK} we define the Lorentz group Racah coefficient by the 
following integral of a product of 6 bulk-to-bulk propagators over 4 copies of $H_3$ :
\be\label{6j}
(6\Delta)=\int_{H_{3}} d\xi_1 \ldots d\xi_4 \prod_{i<j} K_{\Delta_{ij}}(\xi_i,\xi_j).
\ee
Here $i,j=1,\ldots,4$ enumerate the points integrated over, and $\Delta_{ij}$ are the 6 representations that
the $(6\Delta)$ symbol depends on. As an analysis shows, the 
above multiple integral over the hyperbolic space converges only if $\Delta_i$ satisfy certain conditions. In particular,
the integral does converge (after certain gauge fixing that removes the infinite volume of the group) 
for all representations being those of type I. Thus, in the rest of the paper we assume that all representations are of 
type I. Even though
we continue to label representations by their conformal dimensions \eqref{type-1} is implied everywhere.

The above expression \eqref{6j} for the Racah coefficients is our starting point. The formula \eqref{bulk-prop} for the bulk-to-bulk
propagator in terms of two bulk-to-boundary ones gives a way to rewrite \eqref{6j}. Thus, we replace each bulk-to-bulk
propagator as in \eqref{bulk-prop} by two bulk-to-boundary ones. This is achieved by taking to the boundary the middle
point of every edge $ij$. We obtain:
\be\label{6j-b}
(6\Delta)=\int_{S^2} \prod_{i<j} d^2x_{ij} \int_{H_{3}} d\xi_1 \ldots d\xi_4 \prod_{i<j} K_{\Delta_{ij}}(\xi_i,x_{ij})
K_{\bar{\Delta}_{ij}}(x_{ij},\xi_j).
\ee
This formula is a good starting point for computing the Racah coefficients, see \cite{LK}.

In this paper we would like to propose another expression for the Racah coefficient. It arises from a different
way to compute the bulk-to-bulk propagator. Namely, as we shall show in section \ref{sec:new-rep}, 
the bulk-to-bulk propagator can be obtained as an integral over {\it two} boundary points of a product of two bulk-to-boundary 
and one boundary-to-boundary
propagator. We have not yet introduced a boundary-to-boundary propagator. It is defined as the following
function of two boundary points:
\be
\label{2-point}
K_\Delta(x,y) = \frac{1}{|x-y|^{2\Delta}}.
\ee
The reader will recognize in this expression the CFT 2-point function. As will be demonstrated in section \ref{sec:new-rep},
the bulk-to-bulk propagator can be obtained from two bulk-to-boundary and one boundary-to-boundary propagators as:
\be\label{new-rep}
K_{\Delta}(\xi_1,\xi_2) = \frac{1-\Delta}{\pi} \int_{S^2} d^2x d^2y\,\, \frac{K_\Delta(\xi_1,x) K_{\Delta}(\xi_2,y)}{|x-y|^{2\bar{\Delta}}}.
\ee
Thus, instead of taking one middle point to the boundary, in this new representation two points as well as the whole segment connecting
them is taken to the boundary. Using this representation of the bulk-to-bulk propagator, 
we obtain the following expression for the $(6\Delta)$-symbol:
\be\label{6j-new}
(6\Delta)=\frac{\prod_{i<j} (1-\Delta_{ij})}{\pi^6} \int_{S^2} \prod_{i<j} d^2x_{ij} d^2y_{ij} \times \\ \nonumber 
\int_{H_{3}} d\xi_1 \ldots d\xi_4 \prod_{i<j} 
\frac{K_{\Delta_{ij}}(\xi_i,x_{ij}) K_{\Delta_{ij}}(\xi_j,y_{ij})}{|x_{ij}-y_{ij}|^{2\bar{\Delta}_{ij}}}.
\ee

Now let us note that each bulk point is shared exactly by 3 bulk-to-boundary propagators. Let us introduce the following quantity:
\be\label{3-point}
C_{\Delta_1,\Delta_2,\Delta_3}(x_1,x_2,x_3) = \int_{H_{3}} d\xi K_{\Delta_1}(\xi,x_1) K_{\Delta_2}(\xi,x_2) K_{\Delta_3}(\xi,x_3).
\ee
This quantity is nothing but the Clebsch-Gordan coefficient. The
$(6\Delta)$-symbol can be rewritten in terms of these coefficients. To this end, let us introduce the following convenient
notation: $X^i_j=x_{ij}, X^j_i=y_{ij}$. The meaning of these notation is as follows: 
the upper index stands for vertex $i$ which the
point $X^i_j$ on the boundary is closer to, while 
the lower index shows which edge ${ij}$ the point lies on. Using this notation,
we get the following expression for the Racah in terms of the Clebsch-Gordan coefficients:
\be
\label{6j-new-1}
(6\Delta)=\frac{\prod_{i<j} (1-\Delta_{ij})}{\pi^6} \int_{S^2} \prod_{i<j} \frac{d^2X^i_j d^2X^j_i}{|X^i_j-X^j_i|^{2\bar{\Delta}_{ij}}} 
\prod_i C_{\Delta_{ij},\Delta_{ik},\Delta_{il}}(X^i_j,X^i_k,X^i_l),
\ee
where $i\not=j\not=k\not=l$ in the last product.

\begin{figure}\label{fig:ext-tet}
\centering
\epsfig{figure=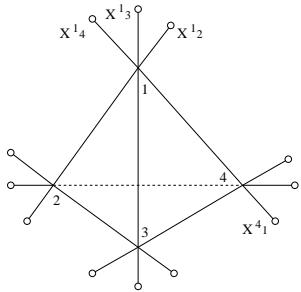, height=1.5in}
\caption{A configuration that dominates the boundary integrations is that
of points where the extended edges meet the boundary. End points of the extended edges
are marked by circles and are at infinity of the hyperbolic space.}
\end{figure}

As
is demonstrated in section \ref{sec:3j}, the bulk integral in \eqref{3-point} can be taken with the result being:
\be\label{3-point-res}
C_{\Delta_1,\Delta_2,\Delta_3}(x_1,x_2,x_3)=\frac{C(\Delta_1,\Delta_2,\Delta_3)}{(x_{12})^{\Delta_1+\Delta_2-\Delta_3} 
(x_{13})^{\Delta_1+\Delta_3-\Delta_2} (x_{23})^{\Delta_2+\Delta_3-\Delta_1}},
\ee
where 
\be\label{C-res}
&{}& C(\Delta_1,\Delta_2,\Delta_3) = \\ \nonumber &{}& \frac{\pi \Gamma(\frac{\Delta_1+\Delta_2+\Delta_3-2}{2})
\Gamma(\frac{\Delta_2+\Delta_3-\Delta_1}{2}) \Gamma(\frac{\Delta_1+\Delta_3-\Delta_2}{2}) 
\Gamma(\frac{\Delta_1+\Delta_2-\Delta_3}{2})}{2 \Gamma(\Delta_1)\Gamma(\Delta_2)\Gamma(\Delta_3)}.
\ee
Substituting this result into the expression \eqref{6j-new-1}, we obtain our final result:
\be
\label{6j-new-2}
(6\Delta)=\frac{\prod_{i<j} (1-\Delta_{ij})}{\pi^6} 
\int_{S^2} \prod_{i<j} \frac{d^2X^i_j d^2X^j_i}{|X^i_j-X^j_i|^{2\bar{\Delta}_{ij}}} \times \\ \nonumber
\prod_i \frac{C(\Delta_{ij},\Delta_{ik},\Delta_{il})}{|X^i_j-X^i_k|^{\Delta_{ij}+\Delta_{ik}-\Delta_{il}} 
|X^i_k-X^i_l|^{\Delta_{ik}+\Delta_{il}-\Delta_{ij}} |X^i_j-X^i_l|^{\Delta_{ij}+\Delta_{il}-\Delta_{ik}}},
\ee
where again $i\not=j\not=k\not=l$. Thus, in words, the Racah coefficient is given by an integral over 12 points in the
boundary of a product of 6 boundary-to-boundary propagators coming from the original edges $ij$ times 12
boundary-to-boundary propagators (3 for every vertex) coming from the Clebsch-Gordan coefficients. The structure
arising is shown in Fig. \ref{fig:ext-tet}. The obtained expression for the Racah coefficient involves only
boundary-to-boundary propagators; all bulk integrals have been taken. It is in this sense that \eqref{6j-new-2}
gives a ``holographic'' realization. It is also holographic in the sense that it is obtained
from the original bulk realization \eqref{6j} by pushing all 6 bulk-to-bulk propagators to the boundary. The intertwiner
at each vertex, however, becomes more complicated than in the bulk realization. It is now given by the
Clebsch-Gordan coefficient \eqref{3-point}.
  
We note that the bulk integrations could have been carried out already at the stage \eqref{6j-b}. In that expression
each bulk point is similarly shared by 3 bulk-to-boundary propagators. Integrals can be taken; the resulting expression
involves an integral over 6 boundary point of a product of 4 Clebsch-Gordan coefficients. This realization is also 
``holographic''. It is more useful than \eqref{6j-new-2} for an actual computation of the Racah coefficient, 
see \cite{LK}. On the other hand, the new 
realization \eqref{6j-new-2} we have obtained in this paper admits an interesting geometrical interpretation. 

In order to arrive to a geometric interpretation let us find out which configuration of boundary points
dominates in the integrations in \eqref{6j-new}. Let us fix positions of the bulk points and consider
only the boundary integrations. Let us consider the limit of all the representations becoming
large $\rho\to\infty$. In this case the stationary phase approximation may be used. 
As is shown in section \ref{sec:new-rep},  when applied to the to the two boundary integrals in \eqref{new-rep}
the stationary phase approximation requires the points 
$x,y$ to be those where the geodesic that passes through $\xi_1,\xi_2$ intersects the boundary. 
This means that the dominant contribution to the boundary integrals in \eqref{6j-new} (when all
the bulk points are fixed) is as shown in Figure \ref{fig:ext-tet}. The picture shown is that of
an ``extended'' tetrahedron, where all the tetrahedron edges are extended till they meet the
boundary. Such an extended tetrahedron plays an important role in the determination of
the hyperbolic tetrahedron volume in terms of volumes of 
ideal (with vertexes at infinity) polyhedra in \cite{Volume}.

Thus, the new formula for the $6j$-symbol that we have proposed have a virtue that it contains only boundary
integrations. It also admits an immediate geometrical interpretation in terms of an extended tetrahedron
shown in \ref{fig:ext-tet}. We see that in the ``holographic'' formulation 
each bulk integration has been replaced by 3 boundary integrations. Geodesics starting at
such 3 boundary points meet inside at the tetrahedron vertex. In the semi-classical approximation
of large representations a geometrical tetrahedron inside the hyperbolic space dominates,
but away from this approximation the geodesics no longer meet and vertexes get blurred. 

The rest of the paper is organized as follows. In section \ref{sec:prop} we compute the bulk-to-bulk propagator using
representation \eqref{bulk-prop}. In section \ref{sec:new-rep} we compute the same bulk-to-bulk
propagator using the new representation \eqref{new-rep} and show that the two quantities coincide. 
We also obtain the stationary phase approximation here.
Section \ref{sec:3j} carries out a computation of the Clebsch-Gordan coefficient. We conclude
with a discussion. Appendix contains formulas necessary to compute bulk and boundary integrals.

\section{Computation of the bulk-to-bulk propagator}
\label{sec:prop}

In this section we explicitly compute the bulk-to-bulk propagator using the formula
\eqref{bulk-prop} as our starting point. To compute the boundary integral we shall use 
the Feynman parameterization, see the Appendix. Thus, we have:
\be
K_\Delta(\xi_1,\xi_2)= \frac{(\xi_1^0)^\Delta (\xi_2^0)^{\bar{\Delta}}}{\Gamma(\Delta)\Gamma(\bar{\Delta})} 
\int_{S^2} d^2x \int dt du\,\, t^{\Delta-1} u^{\bar{\Delta}-1} \\ \nonumber
 e^{-t (\xi_1^0)^2 -u (\xi_2^0)^2 - t|\xi_1-x|^2 -u|\xi_2-x|^2}.
\ee
Let us take the integral over $x$. We get:
\be
K_\Delta(\xi_1,\xi_2)= \frac{(\xi_1^0)^\Delta (\xi_2^0)^{\bar{\Delta}}}{\Gamma(\Delta)\Gamma(\bar{\Delta})} 
\int dt du\,\, t^{\Delta-1} u^{\bar{\Delta}-1} 
e^{-t (\xi_1^0)^2 -u (\xi_2^0)^2} \frac{\pi}{t+u} e^{-\frac{t u}{t+u} |\xi_1-\xi_2|^2}.
\ee
We know that the result must be invariant under the action of the Lorentz group.
Therefore, it can only depend on the hyperbolic distance between the points $\xi_1,\xi_2$. We can always
use the Lorentz group action to put these two points so that $\xi_1=\xi_2$. Then the hyperbolic distance is simply:
\be
\xi_1=\xi_2 \rightarrow l=\log{(\xi_1^0/\xi_2^0)}.
\ee
Thus, let us put $\xi_1=\xi_2$. Let us also make a rescaling of all the variables. We get:
\be
K_\Delta(\xi_1,\xi_2)= \frac{\pi \mu^\Delta}{\Gamma(\Delta)\Gamma(\bar{\Delta})}
\int \frac{dt du}{t+u} \,\, t^{\Delta-1} u^{\bar{\Delta}-1} e^{-t \mu^2-u}.
\ee
Here we have introduced: $\mu=e^l$.
The integrals over $t,u$ can be taken. This can be done, for example, by making a change of variables to:
\be\label{change}
\lambda = \frac{t}{t+u}, \qquad dt du = \frac{u d\lambda du}{(1-\lambda)^2}.
\ee
The range of the new variable is $\lambda\in[0,1]$. We get:
\be
K_\Delta(\xi_1,\xi_2)= \frac{\pi \mu^\Delta}{\Gamma(\Delta)\Gamma(\bar{\Delta})}
\int_0^1 \frac{d\lambda}{1-\lambda} \int_0^\infty du\,\,\left(\frac{\lambda u}{1-\lambda}\right)^{\Delta-1} u^{\bar{\Delta}-1} 
e^{-u(\frac{\lambda \mu^2}{1-\lambda}+1)}.
\ee 
The integral over $u$ is now easy to take. Taking into account the fact \eqref{sum}, we get:
\be
K_\Delta(\xi_1,\xi_2)= \frac{\pi \mu^\Delta}{\Gamma(\Delta)\Gamma(\bar{\Delta})}
\int_0^1 \frac{d\lambda}{1-\lambda}\left(\frac{\lambda }{1-\lambda}\right)^{\Delta-1} 
\frac{1}{\frac{\lambda \mu^2}{1-\lambda}+1}.
\ee
The remaining integral can also be taken. After some simple manipulations we get:
\be
K_\Delta(\xi_1,\xi_2)= \frac{\pi}{\Delta-1}
\frac{\sinh{(\Delta-1)l}}{\sinh{l}}.
\ee
For example, for type 1 representations this reduces to:
\be
K_\rho(\xi_1,\xi_2)= \frac{\pi}{\rho}
\frac{\sin{\rho l}}{\sinh{l}}.
\ee

\section{New representation for the bulk-to-bulk propagator}
\label{sec:new-rep}

In this section we propose another way to obtain the bulk-to-bulk propagator. Instead of using a product of
two bulk-to boundary propagators (for a representation and its dual) integrated over a single boundary point, we take two bulk-to-boundary
propagators for {\it one and the same representation}. In the new construction they end at two different boundary
points. Then we connect these two points by the boundary-to-boundary propagator in the dual representation, 
and integrate over the two boundary points. Thus, let us consider the object:
\be\label{prop-2}
\tilde{K}_{\Delta}(\xi_1,\xi_2) = \int_{S^2} d^2x d^2y K_\Delta(\xi_1,x) K_{\Delta}(\xi_2,y) \frac{1}{|x-y|^{2\bar{\Delta}}}.
\ee
We have denoted this new function of two points in $H_3$ by the same letter. Now it is our aim to show that this
new function is proportional to the bulk-to-bulk propagator computed in the previous section. In principle, this
might be expected. Indeed, the quantity \eqref{prop-2} is also invariant under Lorentz group transformations. Similar to 
\eqref{bulk-prop} it satisfies the (massive) Laplace equation with respect to both arguments. Thus, it should be proportional
to \eqref{bulk-prop}. Here we demonstrate this explicitly, and find the proportionality coefficient.

Let us, as before, introduce the Feynman parameters. We then get:
\be
\tilde{K}_{\Delta}(\xi_1,\xi_2) = \frac{(\xi_1^0)^\Delta (\xi_2^0)^\Delta}{\Gamma^2(\Delta)\Gamma(\bar{\Delta})} 
\int_{S^2} d^2x d^2y \int dt du dv\,\, t^{\Delta-1} u^{\Delta-1} v^{\bar{\Delta}-1} \\ \nonumber
e^{-t (\xi_1^0)^2 -u (\xi_2^0)^2 - t|\xi_1-x|^2 -u|\xi_2-y|^2 - v|x-y|^2}.
\ee
Let us first integrate over $x$,then over $y$, and as before, specialize to the case $\xi_1=\xi_2$. We get:
\be
\tilde{K}_{\Delta}(\xi_1,\xi_2) = \frac{\pi^2 (\xi_1^0)^\Delta (\xi_2^0)^\Delta}{\Gamma^2(\Delta)\Gamma(\bar{\Delta})} 
\int dt du dv\,\, \frac{t^{\Delta-1} u^{\Delta-1} v^{\bar{\Delta}-1}}{tu +tv+uv} e^{-t (\xi_1^0)^2 -u (\xi_2^0)^2}.
\ee
As before, let us rescale all the variables. We get:
\be
\tilde{K}_{\Delta}(\xi_1,\xi_2) = \frac{\pi^2 \mu^\Delta}{\Gamma^2(\Delta)\Gamma(\bar{\Delta})} 
\int dt du dv\,\, \frac{t^{\Delta-1} u^{\Delta-1} v^{\bar{\Delta}-1}}{tu +tv+uv} e^{-t \mu^2 -u}.
\ee
Now the integral over $v$ can be easily taken with the result:
\be
\tilde{K}_{\Delta}(\xi_1,\xi_2) = \frac{\pi^2 \mu^\Delta \Gamma(1-\bar{\Delta})}{\Gamma^2(\Delta)} 
\int \frac{dt du}{t+u}\,\, \left( \frac{t u}{t+u}\right)^{\bar{\Delta}-1} t^{\Delta-1} u^{\Delta-1}  e^{-t \mu^2 -u}.
\ee
Now we again use the change of variables \eqref{change} to get:
\be
\tilde{K}_{\Delta}(\xi_1,\xi_2) = \frac{\pi^2 \mu^\Delta \Gamma(1-\bar{\Delta})}{\Gamma^2(\Delta)} 
\int_0^1 \frac{d\lambda}{1-\lambda} \int_0^\infty du\\ \nonumber (\lambda u)^{\bar{\Delta}-1} 
\left(\frac{\lambda u}{1-\lambda}\right)^{\Delta-1} u^{\bar{\Delta}-1} 
e^{-u(\frac{\lambda \mu^2}{1-\lambda}+1)}.
\ee 
Using the condition \eqref{sum}, and taking the integral over $u$ we get:
\be
\tilde{K}_{\Delta}(\xi_1,\xi_2) = \frac{\pi^2 \mu^\Delta \Gamma(1-\bar{\Delta})}{\Gamma(\Delta)} 
\int_0^1 \frac{d\lambda}{(\lambda(x^2-1)+1)^\Delta}.
\ee
The remaining integral can be taken. The result is, after some manipulations:
\be
\tilde{K}_{\Delta}(\xi_1,\xi_2) =-\frac{\pi^2}{(\Delta-1)^2}
\frac{\sinh{(\Delta-1)l}}{\sinh{l}}.
\ee
Thus, we have verified that the two ways of getting the bulk-to-bulk propagator lead to proportional results.

Let us now find which configuration of points dominates the boundary integrals in \eqref{prop-2} in the
limit of large $\rho$. To analyze this question we do not need the Feynman parameterization. Let us consider
the following double integral:
\be
\int dx dy \,\, e^{-\Delta \log{( (\xi_1^0)^2 + |\xi_1-x|^2 )} -
\Delta \log{( (\xi_2^0)^2 + |\xi_2-y|^2 )} - \bar{\Delta} \log{|x-y|^2}}.
\ee
In the limit of large $\rho$ we have a strongly oscillating integrand and usage of the stationary phase
approximation is justified. To find the configuration that dominates let us differentiate the expression
in the exponent with respect to $x,y$. We get the following system of equations:
\be\label{stat-phase}
\frac{x-\xi_1}{(\xi_1^0)^2 + |\xi_1-x|^2 } = \frac{x-y}{|x-y|^2}, \\ \nonumber
\frac{y-\xi_2}{(\xi_2^0)^2 + |\xi_2-y|^2 } = \frac{y-x}{|x-y|^2}.
\ee
It is not hard to check that each of these two equations is a relation between the coordinates $\xi^0,\xi$ of a 
point that lies on a geodesic and the coordinates $x,y$ of its ends. Indeed, geodesics in hyperbolic space
can be parameterized by their end points $x,y$ on the boundary. Then coordinates of any other point
belonging to this geodesic are given by:
\be
\xi^0 = \frac{|x-y|}{2 \cosh{\rho}}, \qquad \xi = \frac{ x e^\rho + y e^{-\rho}}{e^\rho+ e^{-\rho}},
\ee
where $\rho$ is a parameter along the geodesic. It is chosen in such a way that $\xi(-\infty)=y, \xi(+\infty)=x$.
It is straightforward to check that:
\be
\frac{x-\xi}{(\xi^0)^2 + |\xi-x|^2 } = \frac{x-y}{|x-y|^2}
\ee
is satisfied for all geodesic points. Thus points $x,y$ that solve the system \eqref{stat-phase} are
indeed the end points of the geodesic that passes through $\xi_1,\xi_2$.

\section{3-point function}
\label{sec:3j}

In order to compute the Clebsch-Gordan coefficients \eqref{3-point} integral we again 
use the Feynman parameterization \eqref{feyn}. This is done for each of the bulk-to-boundary propagators.
We get:
\be
C_{\Delta_1,\Delta_2,\Delta_3}(x_1,x_2,x_3) = \frac{1}{\Gamma(\Delta_1)}  \frac{1}{\Gamma(\Delta_2)} \frac{1}{\Gamma(\Delta_3)}
\int_0^\infty dt_1 dt_2 dt_3\, t_1^{\Delta_1-1} t_2^{\Delta_2-1} t_3^{\Delta_3-1} \\ \nonumber
\int_0^\infty \frac{d\xi_0}{\xi_0^3}\, \xi_0^{\sum_i \Delta_i} \int_{S^2} d^2\xi \,
e^{-t_1(\xi_0^2+|\xi-x_1|^2)-t_2(\xi_0^2+|\xi-x_2|^2)-t_3(\xi_0^2+|\xi-x_3|^2)}.
\ee
We now use the formulas \eqref{1}, \eqref{2} of the Appendix to get:
\be
C_{\Delta_1,\Delta_2,\Delta_3}(x_1,x_2,x_3) = \frac{\pi \Gamma(\frac{\sum_i \Delta_i-2}{2})}{2 \Gamma(\Delta_1)\Gamma(\Delta_2)\Gamma(\Delta_3)}
\int_0^\infty dt_1 dt_2 dt_3\, t_1^{\Delta_1-1} t_2^{\Delta_2-1} t_3^{\Delta_3-1} \\ \nonumber
(S_t)^{-(\sum_i \Delta_i)/2} e^{-\frac{1}{S_t}(\sum_{i<j} t_i t_j |x_i-x_j|^2)}.
\ee

We now make a series of changes of variables of integration. The first change is:
\be
t_i = (S_t)^{1/2} t_i' = (\sum_i t_i') t_i', \qquad {\rm det}\left( \frac{\partial t_i}{\partial t_j'}\right) = 2 (S_t)^{3/2}.
\ee
Removing the primes, we get:
\be
&{}& C_{\Delta_1,\Delta_2,\Delta_3}(x_1,x_2,x_3) \\ \nonumber &=& 
\frac{\pi  \Gamma(\frac{\sum_i \Delta_i-2}{2})}{\Gamma(\Delta_1)\Gamma(\Delta_2)\Gamma(\Delta_3)}
\int_0^\infty dt_1 dt_2 dt_3\, t_1^{\Delta_1-1} t_2^{\Delta_2-1} t_3^{\Delta_3-1} e^{- \sum_{i<j} t_i t_j x_{ij}^2 }.
\ee
Here we have introduced:
\be
x_{ij}=|x_i-x_j|.
\ee

The second change of variables is:
\be
t_1 t_2 \to \frac{t_1 t_2}{x_{12}^2}, \qquad t_1 t_3 \to \frac{t_1 t_3}{x_{13}^2}, \qquad t_2 t_3 \to \frac{t_2 t_3}{x_{23}^2}.
\ee
It is easy to see that the integral reduces to:
\be
C_{\Delta_1,\Delta_2,\Delta_3}(x_1,x_2,x_3)=
\frac{1}
{(x_{12})^{\Delta_1+\Delta_2-\Delta_3} 
(x_{13})^{\Delta_1+\Delta_3-\Delta_2} (x_{23})^{\Delta_2+\Delta_3-\Delta_1}} \\ \nonumber
\frac{\pi \Gamma(\frac{\sum_i \Delta_i-2}{2})}{\Gamma(\Delta_1)\Gamma(\Delta_2)\Gamma(\Delta_3)}
\int_0^\infty dt_1 dt_2 dt_3\, t_1^{\Delta_1-1} t_2^{\Delta_2-1} t_3^{\Delta_3-1} e^{- \sum_{i<j} t_i t_j  }.
\ee

It is now possible to take the remaining integral in Feynman parameters by the following change of variables:
\be
t_1 t_2 =u_3, \qquad t_1 t_3=u_2, \qquad t_2 t_3 = u_1,
\ee
so that:
\be
t_1^2 = \frac{u_2 u_3}{u_1}, \qquad t_2^2 = \frac{u_1 u_3}{u_2}, \qquad t_3^2 = \frac{u_1 u_2}{u_3}, \qquad 
\left|{\rm det}\left( \frac{\partial t_i}{\partial u_j}\right)\right| = \frac{1}{2\sqrt{u_1 u_2 u_3}}.
\ee
The integral over $t_i$ thus reduces to:
\be
\frac{1}{2}\int_0^\infty du_1 du_2 du_3 \, u_1^{\frac{\Delta_2+\Delta_3-\Delta_1-2}{2}} u_2^{\frac{\Delta_1+\Delta_3-\Delta_2-2}{2}} 
u_3^{\frac{\Delta_1+\Delta_2-\Delta_3-2}{2}} e^{-u_1-u_2-u_3} = \\ \nonumber
\frac{1}{2} \Gamma(\frac{\Delta_2+\Delta_3-\Delta_1}{2}) \Gamma(\frac{\Delta_1+\Delta_3-\Delta_2}{2}) 
\Gamma(\frac{\Delta_1+\Delta_2-\Delta_3}{2}).
\ee
Thus, we get for the Clebsch-Gordan coefficients:
\be\label{3j}
C_{\Delta_1,\Delta_2,\Delta_3}(x_1,x_2,x_3)=\frac{C(\Delta_1,\Delta_2,\Delta_3)}{(x_{12})^{\Delta_1+\Delta_2-\Delta_3} 
(x_{13})^{\Delta_1+\Delta_3-\Delta_2} (x_{23})^{\Delta_2+\Delta_3-\Delta_1}},
\ee
where 
\be\label{C}
&{}& C(\Delta_1,\Delta_2,\Delta_3) = \\ \nonumber &{}& \frac{\pi \Gamma(\frac{\Delta_1+\Delta_2+\Delta_3-2}{2})
\Gamma(\frac{\Delta_2+\Delta_3-\Delta_1}{2}) \Gamma(\frac{\Delta_1+\Delta_3-\Delta_2}{2}) 
\Gamma(\frac{\Delta_1+\Delta_2-\Delta_3}{2})}{2 \Gamma(\Delta_1)\Gamma(\Delta_2)\Gamma(\Delta_3)}.
\ee
From the CFT point of view, the quantities $C(\Delta_1,\Delta_2,\Delta_3)$ are the
structure constants.

\section{Discussion}

\begin{figure}\label{fig:trun-tet}
\centering
\epsfig{figure=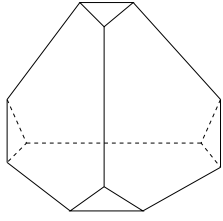, height=1.1in}
\caption{A truncated or {\it stunted} tetrahedron.}
\end{figure}

Few words are in order about a potential importance of the obtained expression for the Racah coefficient. As we
have mentioned above, the main application that we have in mind is that to Euclidean 3d gravity with
negative cosmological constant. Hyperbolic manifolds come in two main classes: those of finite and 
infinite hyperbolic volume. The finite volume manifolds can be triangulated using the so-called
ideal (with all vertexes at infinity) tetrahedra, see \cite{Thurston}, Chapter 3 for more details.
The case of infinite volume manifolds is more
complicated. In this case one can define the so-called convex core, which is a finite volume 
sub-manifold with a totally geodesic boundary. Convex cores can be glued out of the so-called
truncated, or, as Thurston \cite{Thurston} calls them {\it stunted} tetrahedra. An example is shown in 
Figure \ref{fig:trun-tet}. A truncated tetrahedron can also be described as a tetrahedron with
all its vertexes lying {\it outside} of the hyperbolic space. 
An example of a genus 2 handle-body with a totally geodesic boundary obtained
by gluing two truncated tetrahedra is given in \cite{Thurston} Chapter 3.2. 

In order to construct the quantum gravity partition function
for hyperbolic manifolds with boundary it seems necessary to find a $6j$-symbol that would describe
a truncated tetrahedron. By analogy with \eqref{6j} one might expect that such a $6j$-symbol
is given by a multiple integral over positions of vertexes, which would now lie outside
of the hyperbolic space. The space outside of the unit sphere is actually the De-Sitter space,
see e.g. \cite{Gelfand} Chapter V for more details on this. Thus, one would expect to have
to perform an integral over 4 copies of the De-Sitter space. However, unlike the case of
a usual hyperbolic tetrahedron, the integration range for each variable is now expected to 
be a complicated function of all the other variables. Indeed, one should exclude from the
integration configurations in which the tetrahedron lies completely inside the De-Sitter
space (its edges do not intersect the boundary). Indeed, such configurations correspond
to De-Sitter space tetrahedra, not to truncated hyperbolic ones. We thus need a constraint that would
require that the edges do intersect the boundary. However, instead of adding such constraints
it seems to be more practical to just integrate over positions where the intersections happen.
The expression \eqref{6j-new} we have proposed in this paper is exactly of such type. 
Thus, results of the present paper give hope that $6j$-symbols describing the 
truncated hyperbolic tetrahedra can be computed as a multiple integral over
the boundary. Unlike the usual tetrahedron case where a sufficiently simple
bulk realization \eqref{6j} is also known, we do not expect any simple bulk
realization in the truncated case. Thus, one seems to be forced to look for a 
pure boundary, holographic realization. We expect that such realization will be
of a very similar form to \eqref{6j-new}, with boundary-to-boundary propagators
appropriately modified. We hope to report on this in future publications.

It would be interesting to compare the ``holography'' that played role in this paper with the
more conventional 3D holography, for example the one that has been looked at in 
\cite{O'Loughlin:2000wm,Arcioni:2001ds}. The boundary projections that appear in that
context seem vaguely similar. It would be of interest to find direct links, if any. We
shall not address this question in the present work.

\section*{Acknowledgments}

I thank John Barrett and Jorma Louko for important discussions. 
The author was supported by an EPSRC Advanced Fellowship.

\section{Appendix}

Here we give the formulas that are central to the methods of integration that we use. The same method was exploited
in the context of AdS/CFT correspondence, see, for example, \cite{Gleb}. 
The standard Feynman parameter method is based on the following representation:
\be\label{feyn}
\frac{1}{z^\lambda}=\frac{1}{\Gamma(\lambda)} \int_0^\infty dt t^{\lambda-1} e^{-t z}.
\ee

We shall also need the following two integrals:
\be\label{1}
\int_0^\infty \frac{d\xi_0}{\xi_0^3}\, \xi_0^{\sum_i \Delta_i} e^{-\sum_i t_i \xi_0^2} = 
\frac{1}{2} \left( S_t \right)^{1-(\sum_i \Delta_i)/2} \Gamma((\sum_i \Delta_i)/2-1),
\ee
and 
\be\label{2}
\int_{S^2} d^2x e^{-\sum_i t_i |x-x_i|^2} = \frac{\pi}{S_t} e^{-\frac{1}{S_t}(\sum_{i<j} t_i t_j |x_i-x_j|^2)}.
\ee
In both of these formulas:
\be
S_t = \sum_i t_i.
\ee


\begin{thebibliography}{99}

\bibitem{PR} G. Ponzano and T. Regge, ``Semiclassical limit of Racah coefficients'' 
in Spectroscopic and group theoretical methods in physics (Bloch ed.), North-Holland, 1968.

\bibitem{Wor} S. L. Woronowicz and S. Zakrzewski, ``Quantum deformations of the Lorentz group. The Hopf *-algebra level'', 
Compositio Math. {\bf 90} 211--243 (1994)

\bibitem{Gelfand} I. M. Gelfand, M. I. Graev and N. Ya. Vilenkin, ``Generalized Functions'', Vol. 5 ``Integral geometry and representation theory'',
Academic Press, 1966.

\bibitem{Witten} E. Witten, ``Anti De Sitter Space And Holography'', Adv. Theor. Math. Phys. {\bf 2}  253-291 (1998).

\bibitem{BC} J. Barrett and L. Crane, ``A Lorentzian signature model for quantum general relativity'', Class. Quant. Grav. {\bf 17}
3101-3118 (2000).

\bibitem{FK} L. Freidel and K. Krasnov, ``Simple Spin Networks as Feynman Graphs'', J. Math. Phys. {\bf 41}  1681-1690 (2000).

\bibitem{LK} K. Krasnov and J. Louko, ``Racah-Wigner coefficients for Lorentz group in any dimension'', arXiv: math-ph/0502017.

\bibitem{Volume} Y. Cho and H. Kim, ``On the volume formula for hyperbolic tetrahedra'', Discrete Comput. Geom. {\bf 22} 
347-366 (1999).

\bibitem{Thurston} W. Thurston, The geometry and topology of 3-manifolds, 1980 Princeton University notes, available at: 
http://www.msri.org/publications/books/gt3m/

\bibitem{Gleb} G. Arutyunov, S. Frolov and A. C. Petkou, ``Operator Product Expansion of the Lowest Weight CPOs in N=4 SYM${}_4$ at Strong Coupling'',
Nucl. Phys. {\bf B586} 547-588 (2000); Erratum-ibid. {\bf B609} 539 (2001). 


\bibitem{O'Loughlin:2000wm}
  M.~O'Loughlin,
  ``Boundary actions in Ponzano-Regge discretization, quantum groups and
  AdS(3),''
  Adv.\ Theor.\ Math.\ Phys.\  {\bf 6}, 795 (2003).
  
\bibitem{Arcioni:2001ds}
  G.~Arcioni, M.~Carfora, A.~Marzuoli and M.~O'Loughlin,
  ``Implementing holographic projections in Ponzano-Regge gravity,''
  Nucl.\ Phys.\ B {\bf 619}, 690 (2001).
 
\end{thebibliography}
\end{document}